\documentclass[12pt]{article}

\pdfoutput=1

\usepackage{float, natbib, amsmath, amsfonts}
\usepackage{graphicx, rotating, booktabs, array}
\usepackage[margin={1in,1in},includefoot]{geometry}
\usepackage{authblk}
\usepackage{hyperref}
\usepackage{cleveref}
\usepackage{caption}
\usepackage{subcaption}
\usepackage{floatpag}
\usepackage{enumitem}

\newcommand{\abs}[1]{|#1|}
\graphicspath{ {images/} }
\hypersetup{breaklinks,colorlinks,allcolors=blue,pdfpagemode=UseNone}
\author[1]{Nicholas J. Seewald}
\author[1]{Ji Sun}
\author[1]{Peng Liao}

\affil[1]{Department of Statistics, University of Michigan}

\date{}

\title{MRT-SS Calculator: An R Shiny Application for Sample Size Calculation in Micro-Randomized Trials}

\begin{document}
	
	\maketitle
	
	\begin{abstract}
		The micro-randomized trial (MRT) is a new experimental design which allows for the investigation of the proximal effects of a ``just-in-time'' treatment, often provided via a mobile device as part of a mobile health intervention. As with a traditional randomized controlled trial, computing the minimum required sample size to achieve a desired power is a crucial step in designing an MRT. We present MRT-SS Calculator, an online sample-size calculator for micro-randomized trials, built with R Shiny. MRT-SS Calculator requires specification of time-varying patterns for the  proximal treatment effect and expected treatment availability. We illustrate the implementation of MRT-SS Calculator using a mobile health trial, HeartSteps. The application can be accessed from \url{https://pengliao.shinyapps.io/mrt-calculator}.
	\end{abstract}

\section[Introduction]{Introduction}

Due to recent advances in mobile technologies, including smartphones and sophisticated wearable sensors, mobile health (mHealth) technologies are drawing much attention in the behavioral health community. In ``just-in-time'' mobile interventions, treatments,  provided via a mobile or wearable device, are intended to help an individual in the moment. For example, treatments might promote engaging in a healthy behavior when an opportunity arises, or successfully coping with a stressful event. A challenge in developing just-in-time mobile interventions is the limited experimental methodology available to support their development. Current experimental designs do not readily enable researchers to empirically investigate whether a just-in-time treatment had the intended effect, or when and in which context it is useful to deliver a treatment.

Recently, a new experimental design, called the micro-randomized trial (MRT), has been developed to assess the effects of these interventions~\citep{liao2015micro}. In these trials, participants are sequentially randomized between intervention options at each of many occasions (decision times) at which treatment might be provided. 
As with traditional randomized controlled trials, determining sample size is an important part of the process of designing a micro-randomized trial. More specifically, it is important for the scientist to justify the number of participants or experimental units needed in the study to address a specific scientific question with a given power
~\citep{noordzij2010sample}. 

Here, we introduce MRT-SS Calculator, a user-friendly, web-based application designed to facilitate determination of the minimal sample size needed to detect a given effect of a just-in-time intervention. The application was built using R Shiny~\citep{R, shiny}. Commonly, tools used to compute sample sizes can be difficult to use, particularly for non-statisticians. MRT-SS Calculator provides a clean, intuitive user interface which elicits study parameters from the scientist in a thoughtful way, while still offering enough flexibility to accommodate more complex trials. The calculator is based on methodology reviewed in Section~\ref{sec: methods}. In Section~\ref{sec:usage}, we provide a detailed tour of MRT-SS Calculator and describe its use. Finally, in Section~\ref{sec:heartsteps}, we introduce \textit{HeartSteps}, a micro-randomized trial investigating the use of mobile interventions to increase physical activity in sedentary adults. With HeartSteps as an example, we illustrate the use of MRT-SS Calculator in a ``real-world'' setting.

\section[Review of methodology used in MRT-SS Calculator]{Review of methodology used in MRT-SS Calculator} \label{sec: methods}

MRT-SS Calculator implements the methodology developed in~\citet{liao2015micro}. Here, we present a brief review. A micro-randomized trial provides participants with randomly-assigned treatments at each of $T$ decision times, indexed by $t \in [T]$. Depending on the study, the number of decision times $T$ may be in the 100's or 1000's; in the HeartSteps example described in Section~\ref{sec:heartsteps}, $T=210$.   A simplified version of the longitudinal data collected from each participant in an MRT can be written as
\[
	\left\{S_0, I_1, A_1, Y_2, I_2, A_2, Y_3, \ldots, I_t, A_t, Y_{t+1}, \ldots, I_T, A_T, Y_{T+1} \right\},
\]
where $S_{0}$ is a vector of the individual's baseline information (e.g., age, gender).  This calculator is developed for binary treatment, $A_{t}\in\{0,1\}$; that is, there are two intervention options at decision time  $t$.  The proximal response to treatment $A_{t}$ is denoted by  $Y_{t+1}$. $I_{t}$ is an indicator for the participant's ``availability'' at time $t$. At some decision times, the participant may be unavailable for treatment; that is, it may be unethical, scientifically inappropriate, or infeasible to deliver a treatment.   For example, if treatment involves delivering messages  via audible and visual cues on a smartphone, it would be considered unethical to deliver these potentially distracting messages while the participant is driving a car. In such situations, the participant is classified as unavailable, and $I_{t} = 0$.  During time periods when $I_{t} = 0$, participants are not randomized and $A_t$ is left undefined.

At decision time $t$, the proximal effect of a treatment, denoted $\beta(t)$, is defined as
\[
	\beta(t) = E\left[Y_{t+1}\mid I_t = 1,A_t = 1\right] - E\left[Y_{t+1} \mid I_t = 1, A_t = 0\right],
\]
the difference in expected proximal response conditioned on different treatment options. Note $\beta(t)$ is defined only for those participants who are available at decision time $t$. 

We are interested in testing the null hypothesis 
\[
	H_{0}: \beta(t) = 0, \quad t = 1\ldots T
\]
against the alternative
\[
	H_{1}: \beta(t) > 0, \quad \hbox{ for some } t.
\] 
We wish to find the minimal sample size needed to detect $H_{1}$ with desired power. To construct a test statistic and derive a sample size formula, we target alternatives in $H_1$  that are linear in a vector parameter $\beta \in \mathbb{R}^{p}$; in particular, we target $\beta(t)$ of the form
\[
	\beta(t) = Z_t^{\top}\beta, \ t = 1,\ldots,T,
\]
where $Z_{t}$ is a $p \times 1$ vector function of $t$ and covariates that are unaffected by treatment, such as gender, time of day, and day of the week. For example, consider a study  in which $Z_t^{\top}\beta$  is a linear function of  time in days. We refer to this as a ``linear alternative.''   If there are 5 decision times per day, then   $\beta(t) = \beta_{1} + \lfloor \frac{t-1}{5}\rfloor \beta_{2}$, which can be written as $Z_{t}^{\top}\beta$, where $Z_{t}^{T} = \left(1, \lfloor \frac{t-1}{5}\rfloor\right)$ and $\beta = (\beta_{1}, \beta_{2})^{\top}$. Note that $\lfloor \frac{t-1}{5}\rfloor$ translates the index of each treatment occasion to an index of the number of days that have elapsed since the outset of the study.   In Section \ref{sec:alternatives}, we consider other treatment effect trends. 
 
We construct a test statistic based on the least-squares estimator of $\beta$ from the following working model for $E[Y_{t+1}|I_t=1,A_t]$:
\begin{align}
B_t^{\top}\alpha+(A_t-\rho_t)Z_t^{\top}\beta, \quad t = 1, \ldots, T,  \label{working model}
\end{align}
where $B_t$ is a $q \times 1$ vector function of $t$ and covariates that are unaffected by treatment, such as gender, time of day, and day of the week, and $\rho_t = P[A_t = 1]$ is the randomization probability at decision time $t$. The associated test statistic is  given by
$$N \hat{\beta}^{\top} \hat{\Sigma}_{\beta}^{-1} \hat{\beta}$$
where $\hat{\beta}$ is the least-squares estimator that minimizes
\[
\mathbb{P}_{N} \left\{ \sum_{t=1}^TI_t(Y_{t+1} - B_t^{\top}\alpha-(A_t-\rho_t)Z_t^{\top}\beta)^2 \right\},
\]
where $\mathbb{P}_N$ is the average over the sample with size $N$, and $\hat{\Sigma}_{\beta}$ is an estimator of the asymptotic variance of $\sqrt{N}\hat{\beta}$ \citep{liao2015micro}.  The rejection region for the test is 
\begin{align*}
\left\{ N\hat\beta'\hat\Sigma_\beta^{-1}\hat\beta > \frac{N-q-p}{p(N-q-1)} F_{p, N-q-p}^{-1}\left(1-\alpha_0\right)\right\}
\end{align*}
where $F_{p,N-q-p}$ is  the distribution function of a $F$-distribution with $d_1 = p$ and $d_2 = N-q-p$. 

To derive a tractable sample size formula, \citet{liao2015micro} made additional working assumptions. Under these assumptions,  the minimum-required sample size $N$ to detect the alternative with power $1-\beta_{0}$ is found by solving
\begin{align*}
F_{p, N-q-p; c_N}\left(F_{p, N-q-p}^{-1}\left(1-\alpha_0\right)\right)=1-\beta_0
\end{align*}
where $F_{p,N-q-p;c_N}$ is the distribution function of a non-central $F$-distribution with $d_1 = p, d_2 = N-q-p$ and the non-centrality parameter $$c_{N} = Nd^{\top} \left( \sum_{t=1}^{T} E[I_{t}] \rho_{t} (1-\rho_{t}) Z_{t}Z_{t}^{\top}\right)d,$$ 
where $d $ is a $p$-dimensional vector for the standardized treatment effects, i.e., $\beta(t)/\bar \sigma = Z_t^\top d$; see the definition of $\bar \sigma$ in \cite{liao2015micro}.  We note that to calculate the sample size, it is enough to know $q$, i.e., the length of $B_t$ that is used to model the average outcome in (\ref{working model}). In MRT-SS Calculator presented below, we assume $q=3$ for simplicity (e.g., when $B_t$ corresponds to a quadratic pattern). The reader can use the R package (``MRTSampleSize'') to obtain the sample size in the general setting. 

\section[Using MRT-SS Calculator]{Using MRT-SS Calculator} \label{sec:usage}

In this section, we will explain how to use MRT-SS Calculator. In brief, the user should  provide basic information about the study as well as the alternative hypothesis and expected availability of participants over the course of the trial. MRT-SS Calculator will return either the minimum required sample size to achieve a specified power, or the power achievable given a specified number of participants. We explore each of these components in detail below.

\subsection[Study Setup]{Study setup}

Scientists using MRT-SS Calculator are first prompted to provide specific details describing their planned trial. These include the study's duration in days, the daily number of decision times at which treatment is randomly assigned, and the probability of being randomized to receive treatment.

\begin{figure}
	\centering
	\begin{subfigure}[t]{.45\textwidth}
		\includegraphics[width=\textwidth]{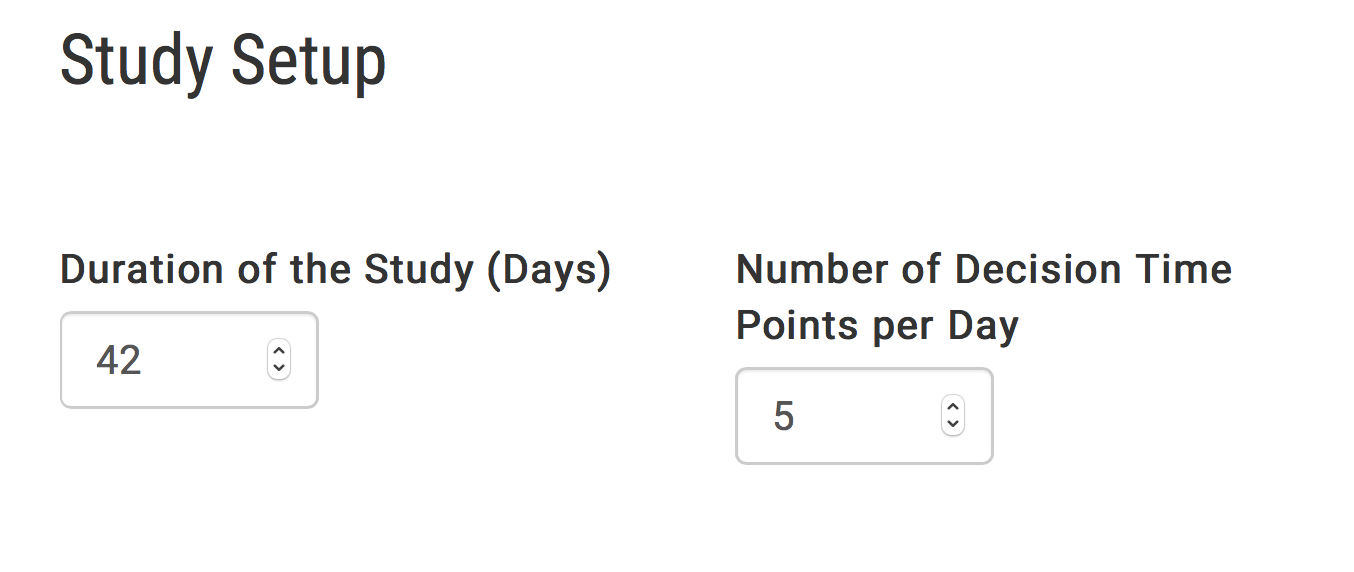}
		\caption{User specification of study duration in days and the number of randomizations per day.}
	\end{subfigure}
	\begin{subfigure}[t]{.45\textwidth}
		\includegraphics[width=\textwidth]{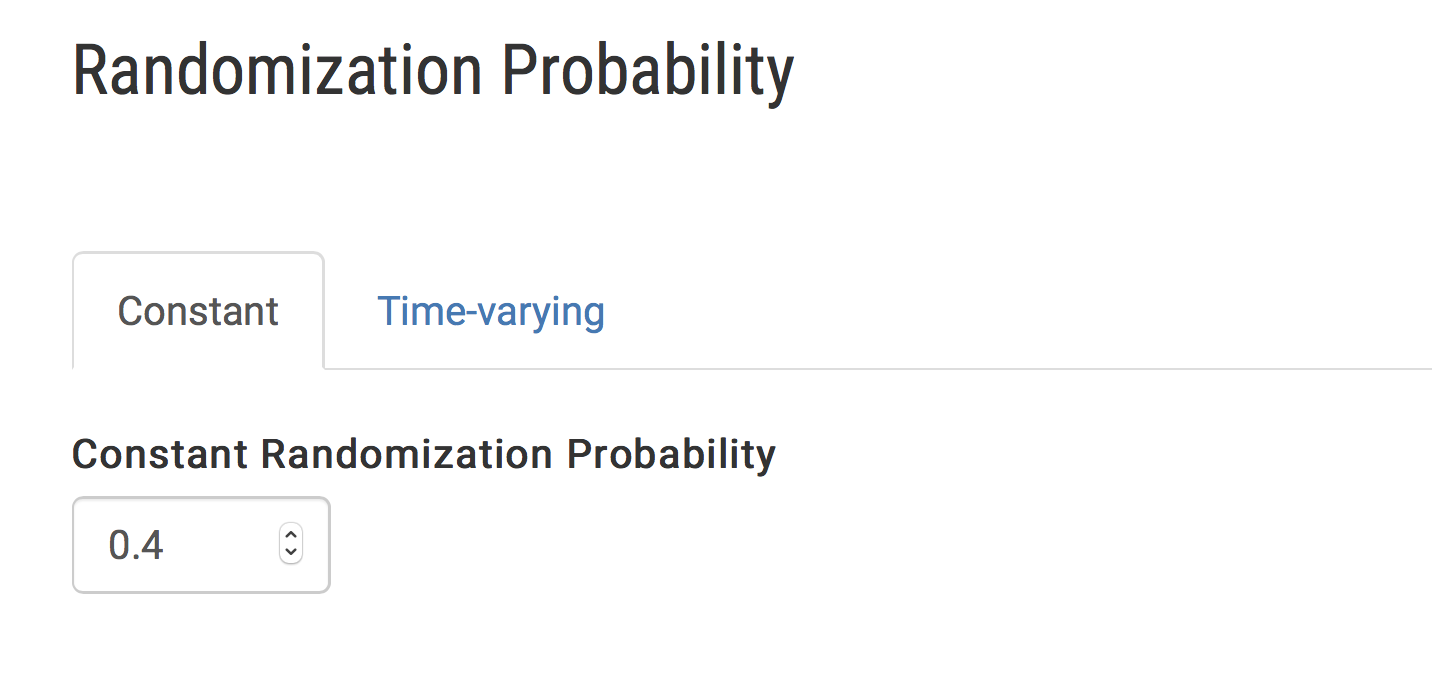}
		\caption{User specification of a constant probability of randomization to treatment.}
	\end{subfigure}
	\caption{Study setup for a 42-day trial in which participants are randomized to receive treatment up to five times per day, conditional on $I_{t} = 1$ (a), each with probability 0.4 (b).}
	\label{fig:setup}
\end{figure}

For example, Figure~\ref{fig:setup} describes a study which takes place over 42 days with up to 5 randomizations per day, where treatment is delivered with probability 0.4 at each decision time ($A_t=1$ if treatment is delivered, $A_t=0$ if no treatment). Using the notation established in Section \ref{sec: methods},  $T = 42 \times 5 = 210$,  and $\rho_{t} = P\left[A_{t} = 1\right] = 0.4$ for all $t\in [T]$. Given this randomization probability, participants in the study will be delivered an average two treatments per day, provided they are available. Treatments are not provided when the participant is unavailable (and thus the participant is not randomized), for the reasons described above.

\begin{figure}
\centering
\includegraphics[width=.75\textwidth]{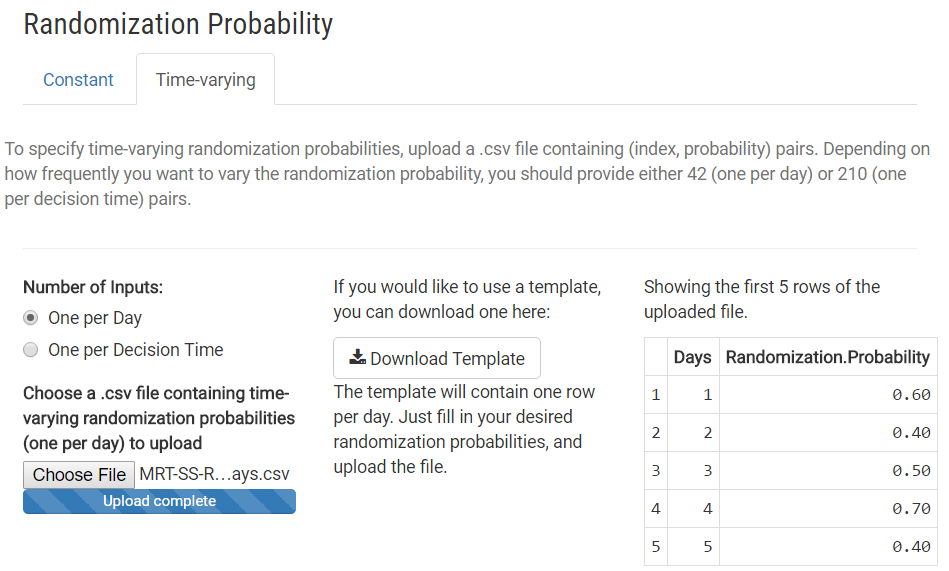}
\caption{Time-varying randomization probabilities. The user specifies whether to provide probabilities which change either daily or at each decision time, and uploads a .csv file containing (index, probability) pairs. A template file is provided for ease of use, and a preview of the uploaded file is shown for verification.}
\label{fig:uploadprobs}
\end{figure}

Some users may wish to vary the randomization probability over the course of the study. MRT-SS Calculator is flexible enough to accommodate this. The user can upload a .csv file containing randomization probabilities for each day of the study, or for every individual decision time point (see Figure \ref{fig:uploadprobs}).

\subsection[Specifying patterns for expected availability]{Specifying patterns for expected availability} \label{sec:avail}

\begin{figure}
	\centering
	\begin{subfigure}{\textwidth}
		\centering
		\includegraphics[height=.25\textheight]{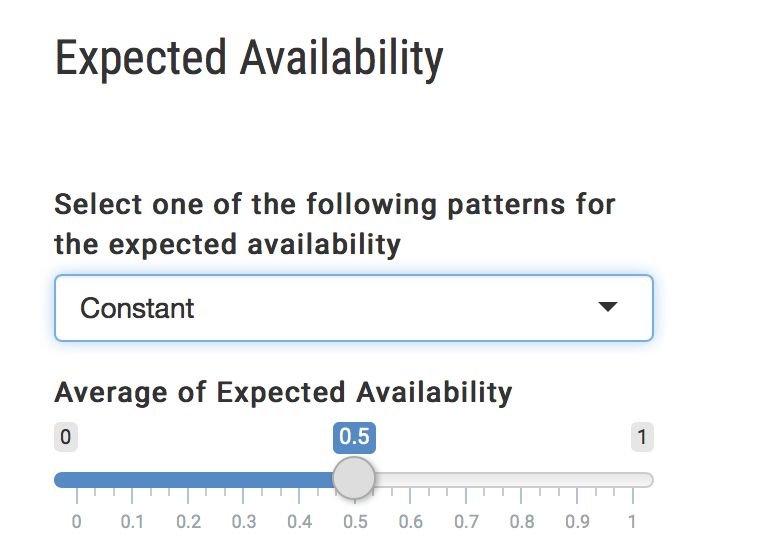}
		\includegraphics[height=.25\textheight]{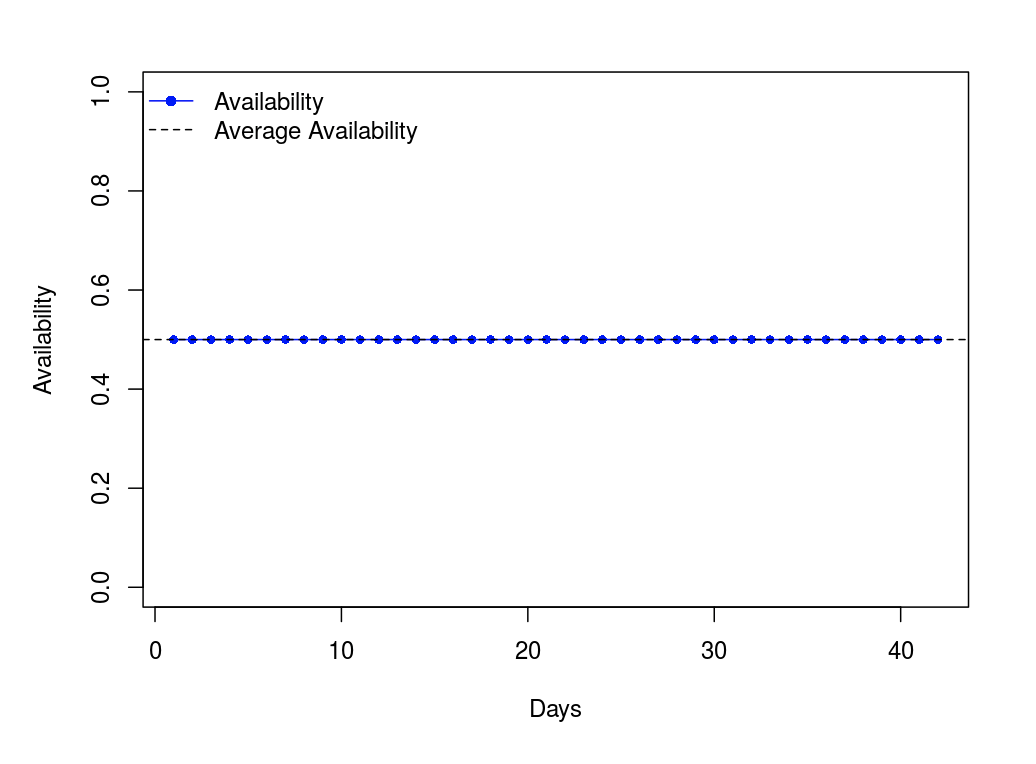}
		\caption{Specification of a constant availability pattern over the course of the trial.}
	\end{subfigure}
	\begin{subfigure}{\textwidth}
		\centering
		\includegraphics[height=.25\textheight]{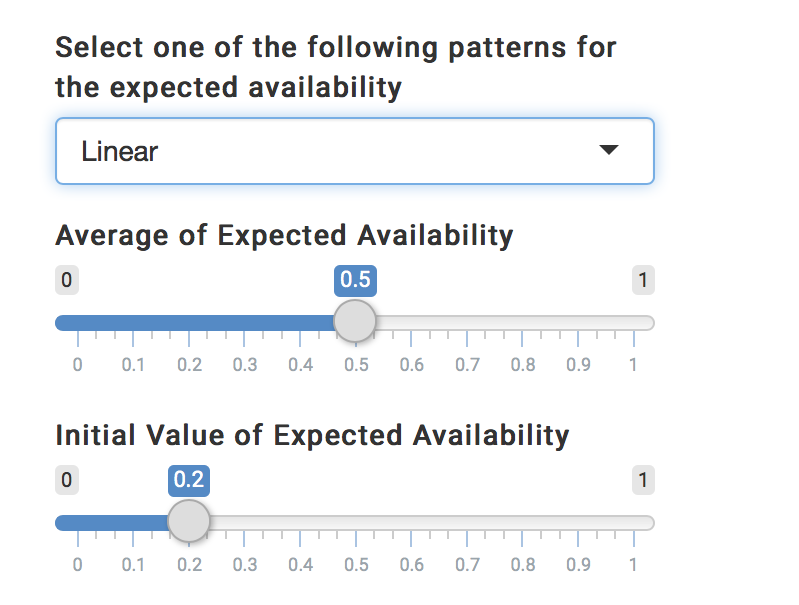}
		\includegraphics[height=.25\textheight]{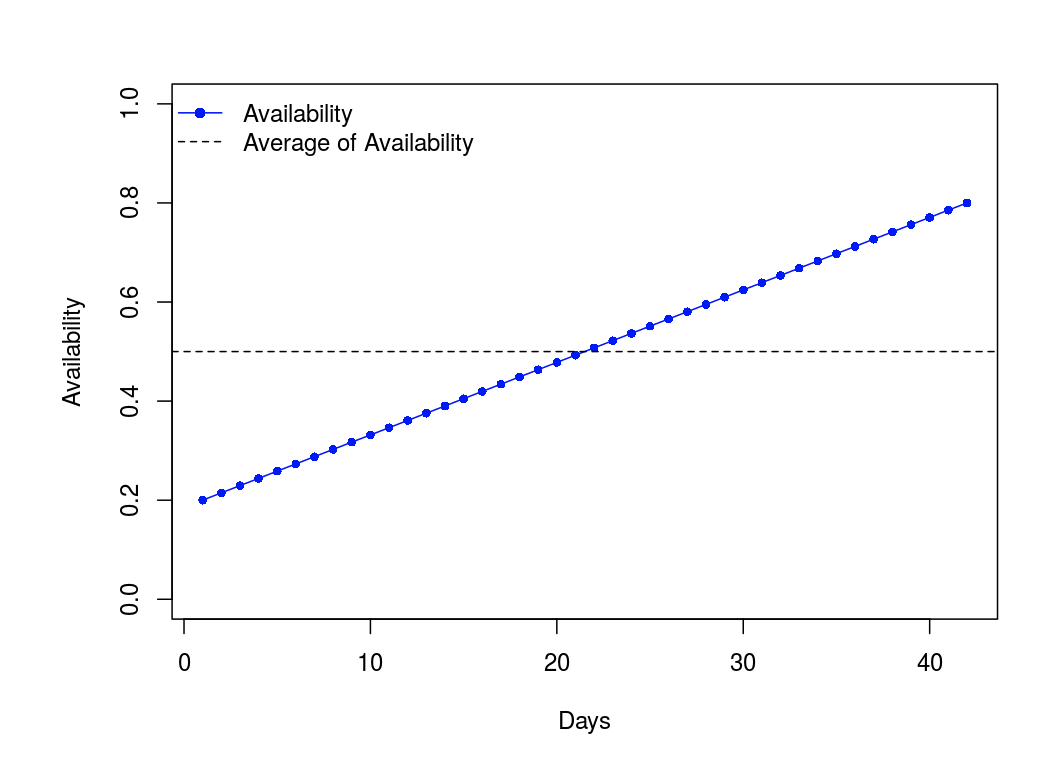}
		\caption{Specification of a linearly-increasing availability pattern over the course of the trial. A linearly-decreasing pattern can be specified by adjusting the initial value and average.}
	\end{subfigure}
	\begin{subfigure}{\textwidth}
		\centering
		\centering
		\includegraphics[height=.28\textheight]{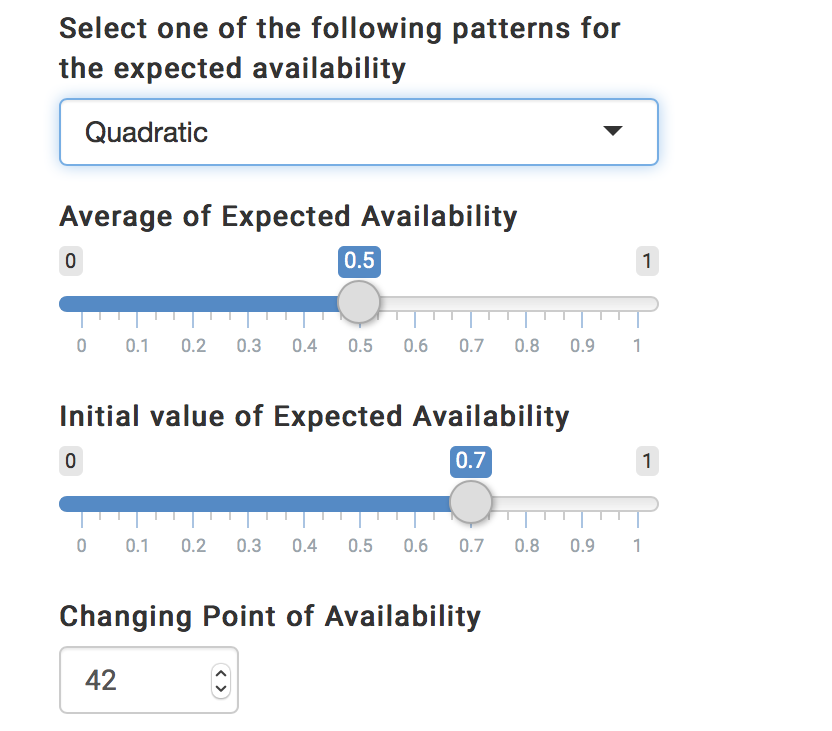}
		\includegraphics[height=.28\textheight]{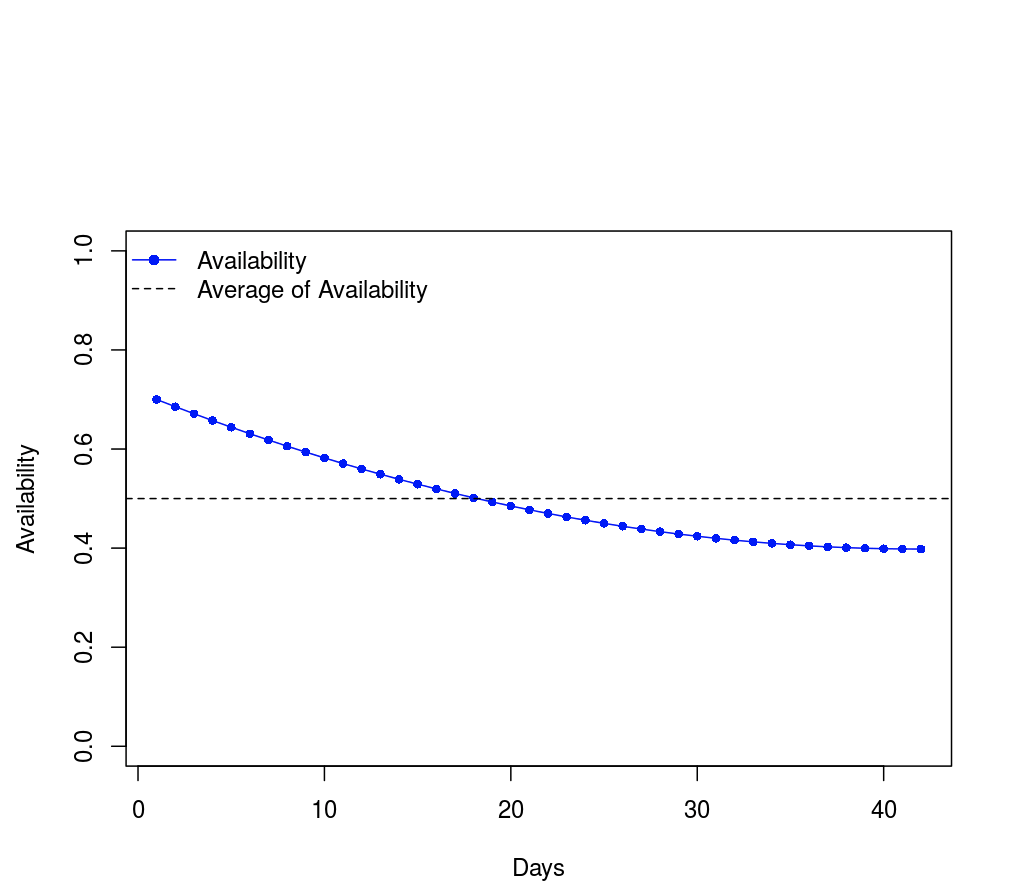}
		\caption{Specification of a quadratic pattern of availability. In the pictured trend, availability decreases over the course of the trial. ``Changing point'' refers to the day on trial at which expected availability is a maximum or minimum. }
	\end{subfigure}
	\caption{Three examples of possible patterns of expected availability, each falling in one of three classes: constant (a), linear (b), and quadratic (c).}
	\label{fig:availability patterns}
\end{figure}

MRT-SS Calculator requires the specification of a time-varying expected availability $E[I_t]$ for each decision time $t \in [T]$. Recall from Section~\ref{sec: methods} that availability may vary according to many factors, including, for example, whether the participant has turned off the intervention. MRT-SS Calculator provides three classes of trends for expected availability: constant, linear, and quadratic (see Figure~\ref{fig:availability patterns}).  These trends are averaged over each day, i.e., if there are multiple decision times per day then the trend is in the average over the multiple decision times.

The different classes of expected availability patterns correspond to a variety of scenarios. For example, if it is believed that participants will be more likely to turn off the intervention as the study goes on, then expected availability will decrease. This might occur if, for example, participants  find the interventions more burdensome as the study goes on.

\begin{figure}
	\centering
	\begin{subfigure}{\textwidth}
		\centering
		\includegraphics[width=0.45\textwidth] {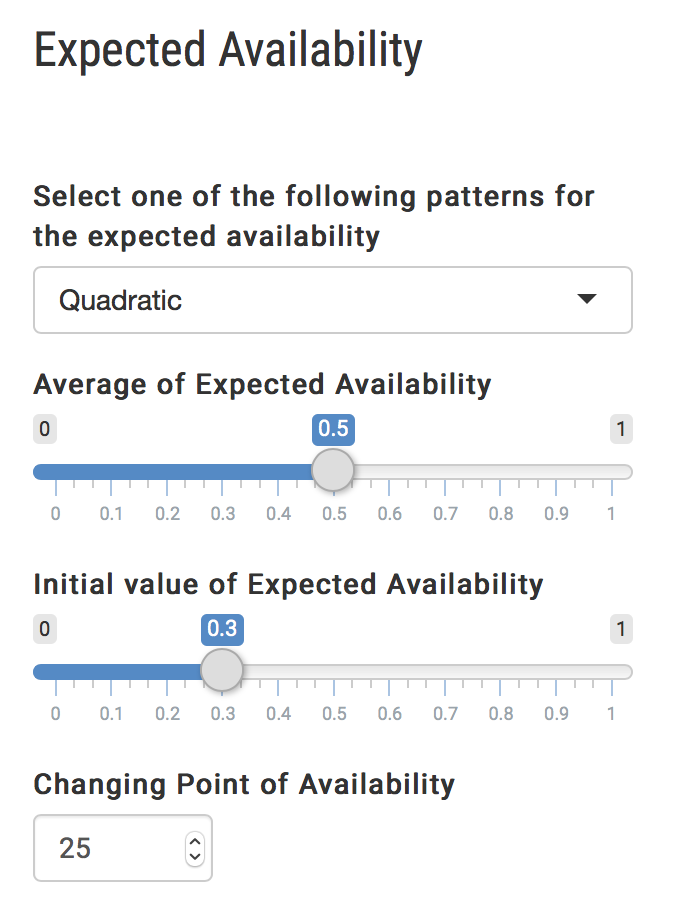}
		\includegraphics[width=0.5\textwidth] {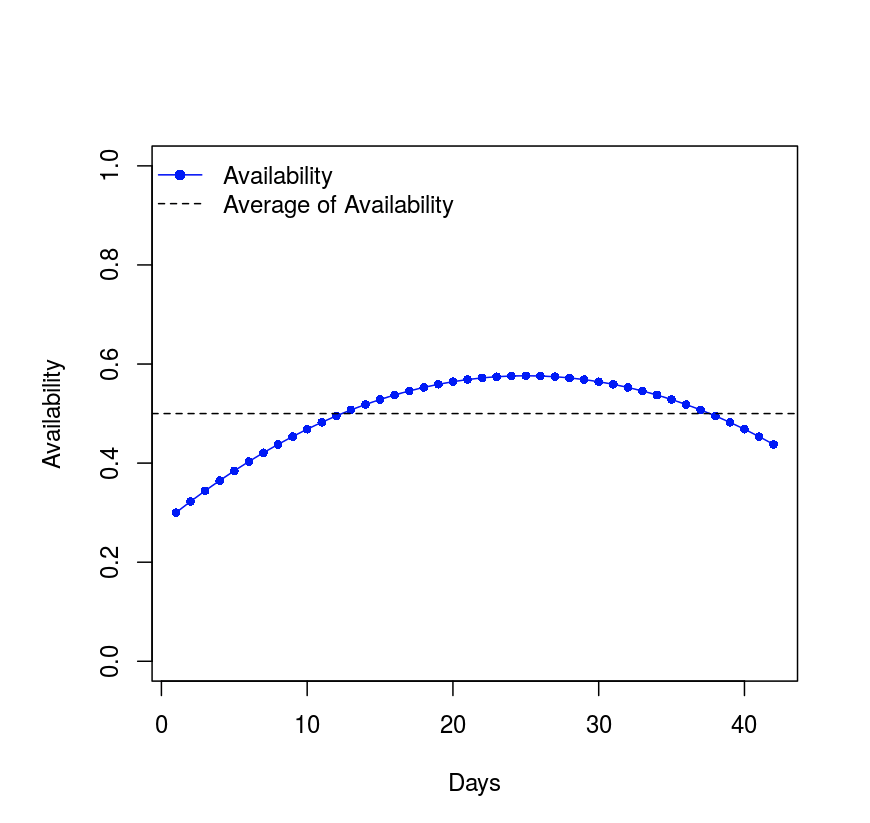}
		\caption{A concave quadratic pattern of expected availability. Availability increases until day 25, when it is maximized, then decreases until the end of the trial.}
	\end{subfigure}
	\begin{subfigure}{\textwidth}
		\centering
		\includegraphics[width=0.45\textwidth] {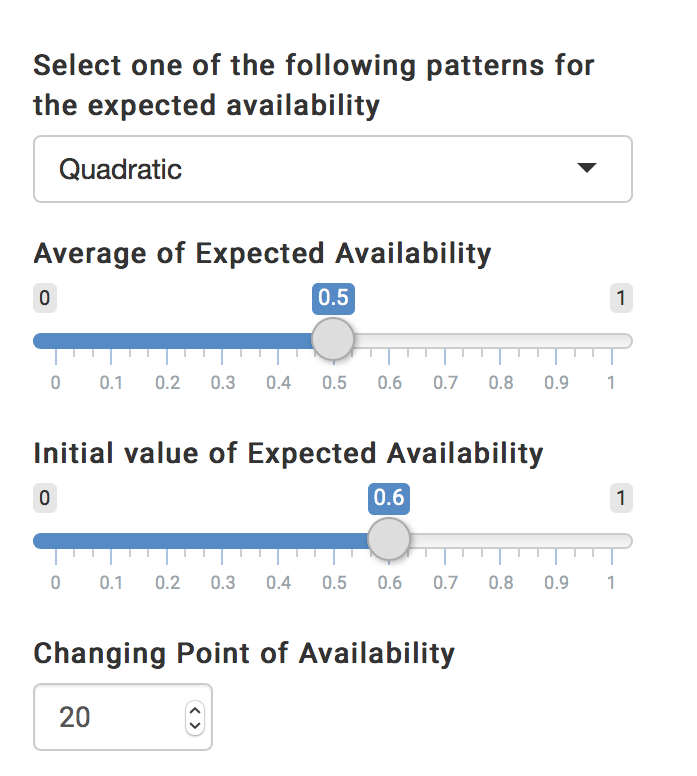}
		\includegraphics[width=0.5\textwidth] {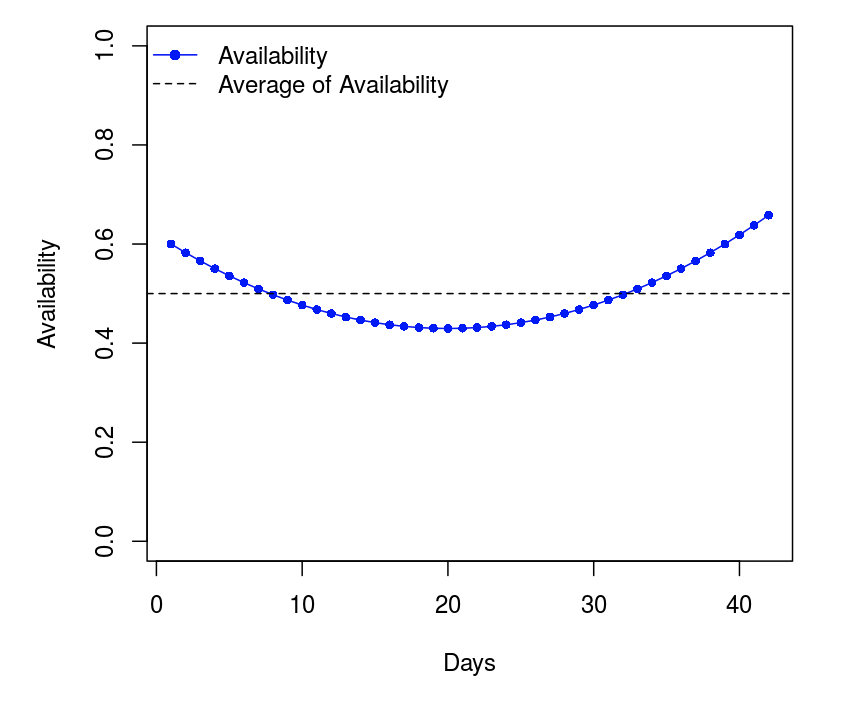}
		\caption{A convex quadratic pattern of expected availability. Availability decreases until day 20, when it is minimized, then increases until the end of the trial.}
	\end{subfigure}
\caption{Different patterns of expected availability in the quadratic class.}
\label{fig:quadratic availability}
\end{figure}

MRT-SS Calculator requires the user to provide inputs which fully specify the pattern of expected availability over the course of the study. For example, after selecting the quadratic class of trends, the user is prompted to provide an estimate of participants' average availability throughout the study, the estimated availability for participants at the outset of the trial, and the day of maximum or minimum availability (the ``changing point''). See Figure~\ref{fig:quadratic availability} for examples of how different values of these inputs change the pattern of expected availability.

\subsection[Specifying  the targeted alternative effect]{Specifying the targeted alternative effect} \label{sec:alternatives}

\begin{figure}
	\centering
	\begin{subfigure}{\textwidth}
		\centering
		\includegraphics[height=.25\textheight]{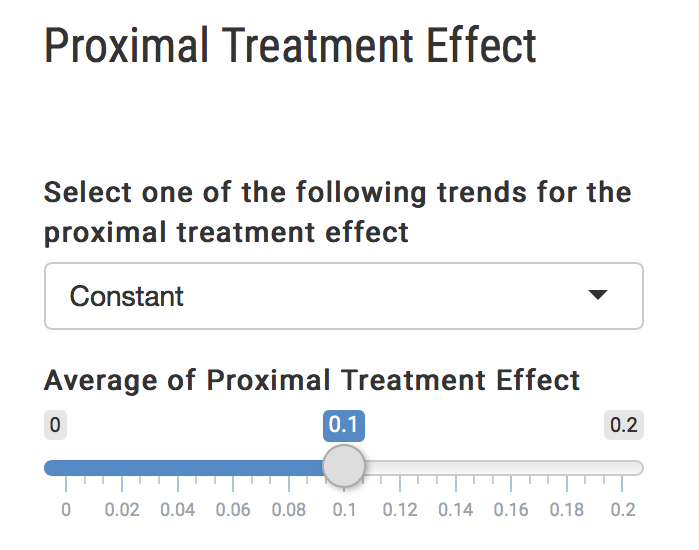}
		\includegraphics[height=.25\textheight]{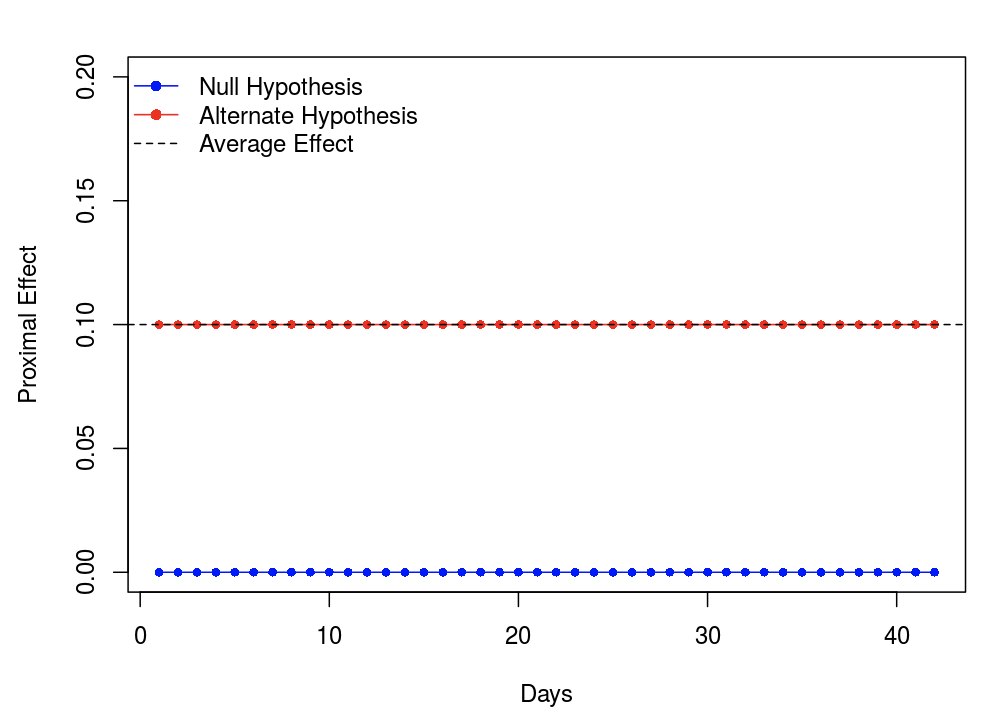}	
		\caption{Specification of a constant targeted alternative proximal treatment effect.}
	\end{subfigure}
	\begin{subfigure}{\textwidth}
		\centering
		\includegraphics[height=.3\textheight]{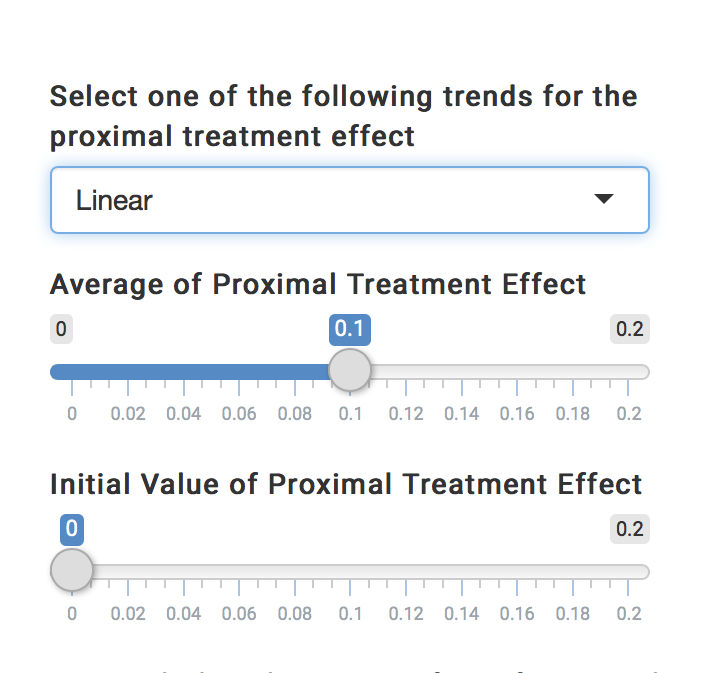}
		\includegraphics[height=.3\textheight]{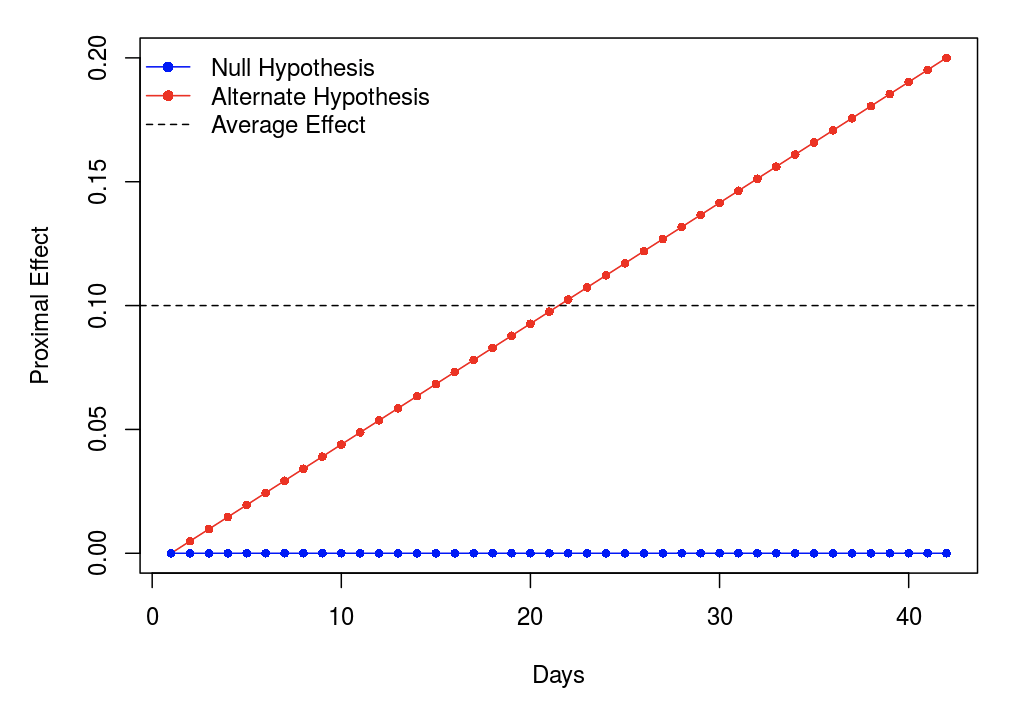}
		\caption{Specification of a linearly-increasing targeted alternative proximal treatment effect. Linearly decreasing effects can be specified by changing the average and initial values.}
	\end{subfigure}
	\begin{subfigure}{\textwidth}
		\centering
		\includegraphics[height=.3\textheight]{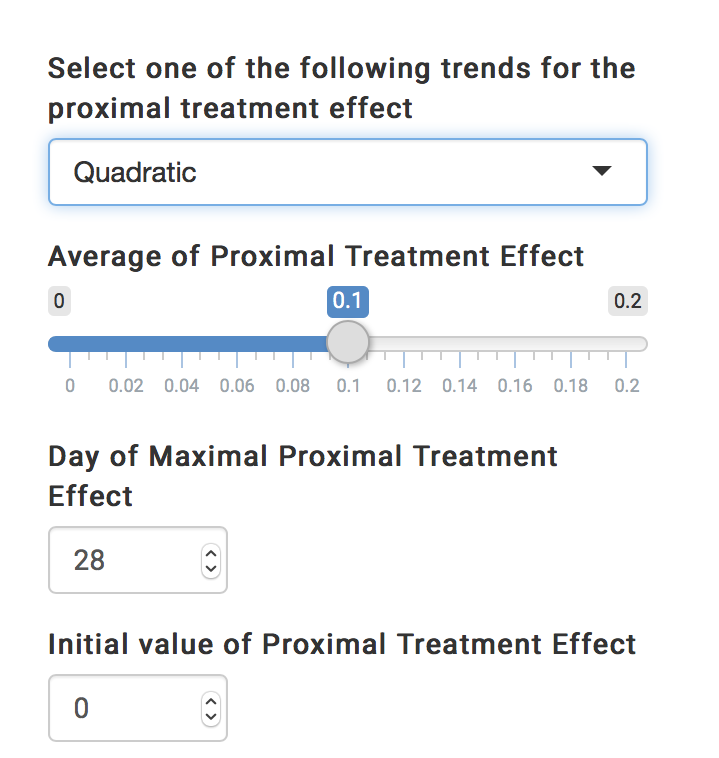}
		\includegraphics[height=.3\textheight]{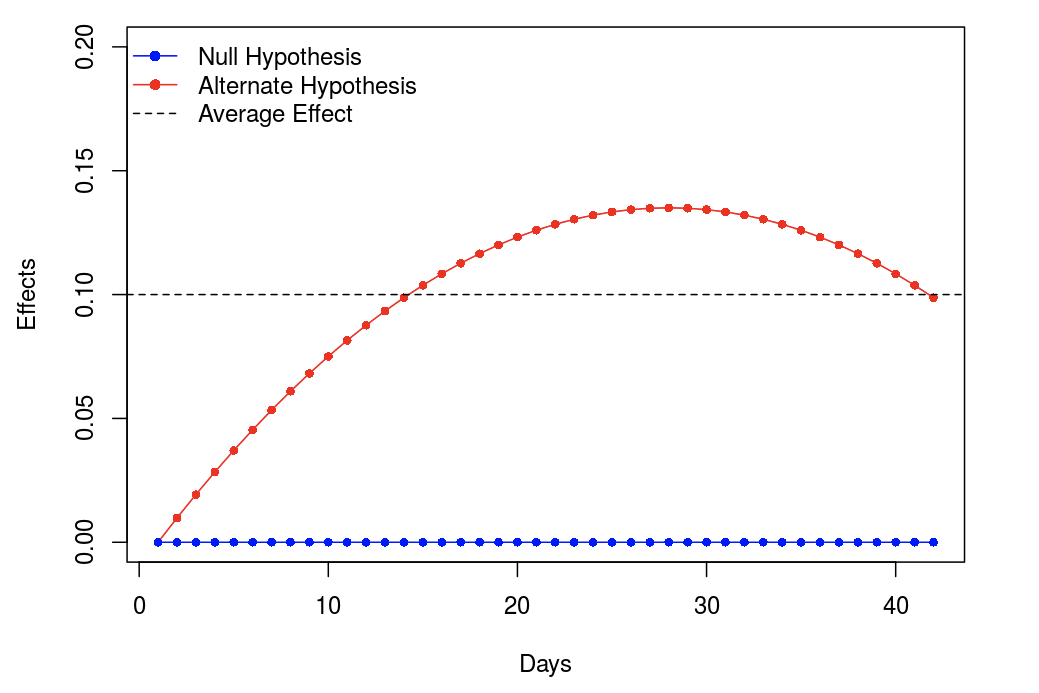}
		\caption{Specification of a quadratic targeted alternative proximal treatment effect.}
	\end{subfigure}
	\caption{Example inputs for the three available classes of trends for the standardized proximal treatment effect: constant (a), linear (b), and quadratic (c). Included in the plots are the null hypothesis ($\beta(t) = 0$ for all $t$) in blue, the specified average treatment effect in black, and the alternative $\beta(t) = Z_{t}^{\top}\beta$ in red.
	}\label{fig:proximal treatment effect}
\end{figure}

MRT-SS Calculator requires the user to specify the desired detectable standardized treatment effect $d$. Recall from Section~\ref{sec: methods} that the targeted  alternative effect is of the form $\beta(t) = Z_{t}^{\top}\beta$, where $Z_{t}$ is some function of time.  Here $d$ is a standardized $\beta$; see \cite{liao2015micro}. MRT-SS Calculator allows the user to choose the form of $Z_{t}$ from constant, linear, or quadratic classes of trends (see Figure \ref{fig:proximal treatment effect}).  In all classes the trends are averaged over each day, as with the pattern of expected availability (see Section~\ref{sec:avail}).

Each of these classes corresponds to a different possible targeted alternative.   A constant trend is most useful if the user believes that the effect of the treatment will be relatively stable over the duration of the study.   A linear trend is useful if, for example, users believe that the effect of the treatment may grow over time and is unlikely to dissipate at later decision times.  A quadratic trend would be useful if it is believed that the treatment effect will grow initially but may dissipate with time as  participants begin to ignore or disengage from treatment.

\subsection[Output]{Output} \label{sec:result}

\begin{figure}
  \centering
  \begin{subfigure}{\textwidth}
  	\centering
  	\includegraphics[width=0.7\textwidth]{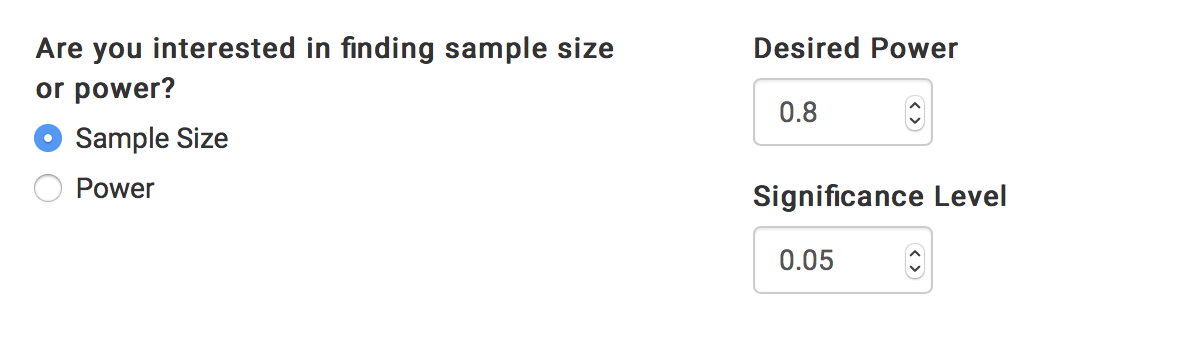}
  	\caption{Selection of calculator output. To determine minimum-required sample size for a trial, the user inputs the desired power to detect the targeted alternative proximal treatment effect, and the significance level for the planned hypothesis test.}
  \end{subfigure}
  \begin{subfigure}{\textwidth}
  	\centering
  	  \includegraphics[width=1\textwidth]{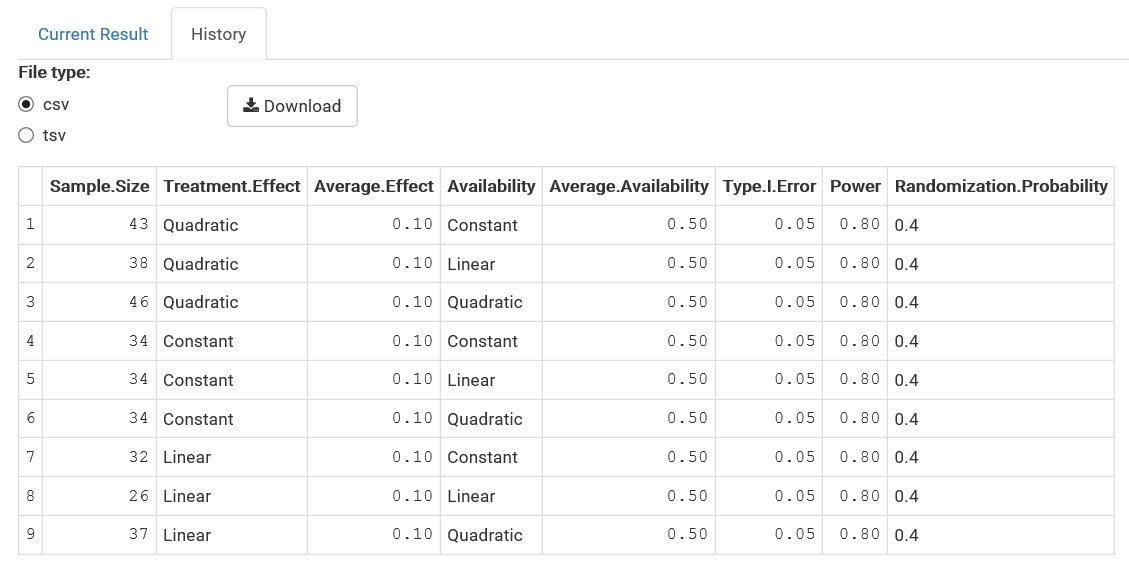}
  	  \caption{Computed minimum-required sample sizes for all combinations of patterns in Figures~\ref{fig:availability patterns} (availability) and \ref{fig:proximal treatment effect} (proximal treatment effect) with study setup as in Figure~\ref{fig:setup}, desired power 0.8 and significance level 0.05.}
  \end{subfigure}
\caption{Output specification and result history. After having provided all required inputs described in Section~\ref{sec:usage}, the user selects the type of outcome desired (a). The user can access a record of past outputs from the current session (b).}
\label{fig:result_samplesize}
\end{figure}

MRT-SS Calculator allows the user to choose to compute either the minimum sample size required to achieve a specified  power or the achievable power given a specified sample size. In both cases, the  significance level of the test must be supplied. Figure \ref{fig:result_samplesize} shows the use of MRT-SS Calculator to obtain a minimum sample size for a 42-day study with 5 decision times per day. All computed results from a session are saved, and users can view and/or download all past output from their current session.

\subsection[Error handling]{Error handling} \label{sec:errors}

\begin{figure}
	\thisfloatpagestyle{empty}
	\begin{subfigure}{0.5\textwidth}
		\centering
		\includegraphics[width=.5\textwidth]{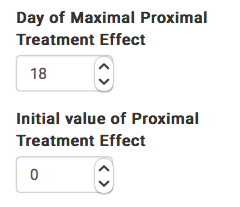}
	\end{subfigure}
	\begin{subfigure}{0.5\textwidth}
		\includegraphics[width=\textwidth]{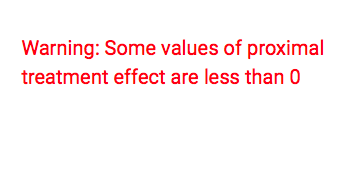}
	\end{subfigure}
	\caption{Error handling for inputs which lead to negative proximal treatment effects. In a 42-day study, choosing ``Day of Maximal Proximal Effect'' less than 22 when using the quadratic class with an initial effect of zero will result in a negative value of proximal treatment effect on some days. The application will produce an error message.}
	\label{fig:warning_maximum}
\end{figure}

The MRT-SS Calculator delivers warnings when inappropriate inputs are provided. For example, the warning provided in Figure~\ref{fig:warning_maximum} will appear if the entered ``Day of Maximal Proximal Effect'' is less than 22  in 42-day study when the initial effect is set to be 0.  This is because some values of the resulting treatment effect are less than 0, which should be avoided. When the calculated sample size is less than 10, MRT-SS Calculator will return 10 with a warning.  In Appendix~\ref{sec:appendix-simulation}, we conduct a simulation study to investigate different scenarios in which the required sample size is less than 10, or equivalently the estimated power for a sample size of 10 is larger than the desired power.  In particular, we compare the power estimated by MRT-SS Calculator with the simulated power under different generative models. In general, the power performance under a variety of generative models when sample size is equal to 10 is slightly degenerated compared to scenarios with relatively large sample sizes.  

\section[An Example: using the MRT-SS Calculator for HeartSteps]{An example: HeartSteps} \label{sec:heartsteps}

\subsection[Introduction to HeartSteps]{Introduction to HeartSteps}

\textit{HeartSteps} is a mobile intervention study designed to increase physical activity for sedentary adults~\citep{heartsteps}. HeartSteps uses a mobile application to investigate the effects of in-the-moment activity suggestions. These are messages which encourage the participant to engage in physical activity and which appear on the lock screen of the participant's mobile phone. The suggestions may vary in content depending on a number of contextual factors such as weather, the participant's location, or the time of day. Participants are randomized to either receive or not receive a suggestion at five pre-specified decision times throughout the day. These time points correspond roughly to a morning commute, mid-day, mid-afternoon, an evening commute, and after dinner.  When a suggestion is delivered, it is displayed on the lock screen of the phone, which then plays a notification sound, vibrates, or lights up. 

In HeartSteps, data are collected both passively via sensors and actively through participant self-report. Each participant is provided a Jawbone UP activity tracker which monitors and records step count. Furthermore, sensors on the phone are used to collect a variety of information at each of the five decision times during the day. This information includes the participant's current location and activity status (e.g., walking, driving, etc.). If sensors indicate that the individual is likely walking or driving a car, activity suggestions are not delivered (the availability indicator $I_t$ is set to 0).

\subsection[Illustration of MRT-SS Calculator with HeartSteps]{Illustration of MRT-SS Calculator with HeartSteps}

HeartSteps is a 42-day trial with five decision times per day, so that $T=210$. Suppose we wish to size the trial to detect a given proximal effect of the intervention on a participant's step count. Notice that this is a binary treatment: participants are randomized to be delivered a suggestion or not. Suggestions are delivered with constant probability $\rho = P[A_{t} = 1] = 0.4$ over the course of the study, so that if a participant is always available, an average of two messages are delivered per day. 

The proximal treatment effect may vary across time for a variety of reasons. In HeartSteps, the treatment  effect might initially increase, as it is believed that participants will enthusiastically engage with the intervention at the outset. Then, as the study goes on, some participants may disengage or begin to ignore the activity suggestions due to habituation, so we expect a decreasing proximal treatment effect. Thus, a plausible target alternative effect would be quadratic in time.

\begin{figure}
	\centering
	\begin{subfigure}{\textwidth}
		\centering
		\includegraphics[width=1\textwidth]{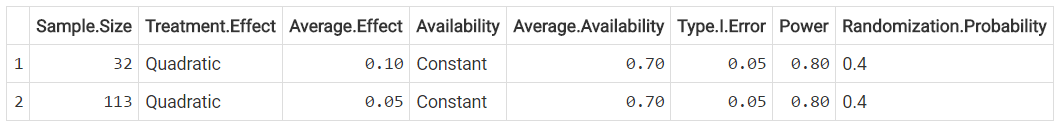}
		\caption{Example sample size output for HeartSteps with a given target power of 80\% and varying target average proximal treatment effects.}
	\end{subfigure}
	\begin{subfigure}{\textwidth}
		\includegraphics[width=1\textwidth]{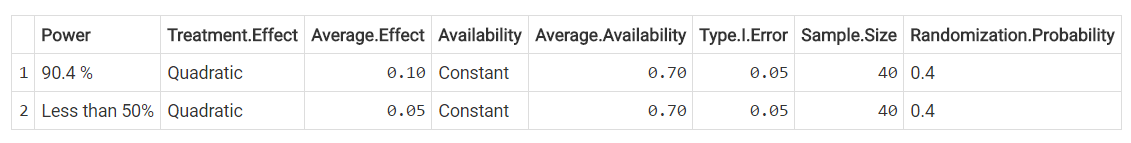}
		\caption{Example power output for HeartSteps with a given sample size of 40 and varying target average proximal treatment effects. The application does not display power less than 50\%.}
	\end{subfigure}
	\caption{Illustrative sample size (a) and power (b) calculations for HeartSteps. The result is displayed in the first column, while the remaining columns are used to describe inputs provided by the user.} 
	\label{fig:ex_sizes}
\end{figure}

For example, we might be interested in the sample size needed to achieve at least 80\% power when there is no initial treatment effect on the first day and the maximal proximal effect comes around day 28, or the amount of power we can guarantee with a sample size of 40.  The results of sample sizes and power calculations for HeartSteps are provided in Figure~\ref{fig:ex_sizes}.

\section[Conclusion]{Conclusion}

MRT-SS Calculator is designed to conduct sample size and power calculations  for micro-randomized trials, a new experimental framework which allows for the investigation of the effects of just-in-time mobile health interventions. Such interventions are of interest not only in the realm of physical activity, as in the example in Section~\ref{sec:heartsteps}, but also in the study of smoking cessation, obesity, congestive heart failure, and healthy eating, among others. MRT-SS Calculator is accessible, designed to elicit trial information from the scientist through a clear, well-explained sequence of inputs and  provides real-time visual feedback. The calculator is also flexible, and capable of sizing trials which vary in complexity. MRT-SS Calculator can be used to quickly and reliably determine the minimal sample size needed to achieve a given power, or the power achievable given a sample size. MRT-SS Calculator can be used by scientists to ease the burden of designing micro-randomized trials to investigate proximal treatment effects.

\section*{Acknowledgments}

The development of this application, as well as the preparation of the manuscript, was undertaken with support from NIAAA R01 AA023187, NIDA P50 DA039838, NIBIB U54EB020404, and NHLBI/NIA R01 HL125440. The authors wish to thank Prof. Susan A. Murphy for her thoughtful advice and support throughout the course of this work.

\bibliography{Seewald-Sun-Liao-MRT-SS-Calculator-ArXiv}

\newpage
\appendix
\renewcommand{\thesection}{\Alph{section}}
\setcounter{table}{0}
\renewcommand{\thetable}{A\arabic{table}}
\setcounter{figure}{0}
\renewcommand{\thefigure}{A\arabic{figure}}
\section{Simulation study}
\label{sec:appendix-simulation}
In this Appendix, we conduct a simulation study to investigate the power performance when sample size is 10 in settings where the theoretical power (with type I error rate $\alpha = 0.05$) is above 0.8; or, equivalently, the required sample size to achieve 0.8 power is below 10. Such situations are likely to occur if: (a) The total number of decision times $T$ is relatively large, due to either a large number of days or large number of decision times per day, (b) the provided average (standardized) proximal treatment effect is relatively large (say, 0.15), or the parameterization of the treatment effect is relatively simple, e.g. constant or linear, and (c) the average of expected availability throughout the study is relatively large. In the following, we provide simulation results for the above scenarios under different generative models.

First, we consider the case when the working assumptions made to obtain a tractable sample size calculation are satisfied; see \citet{liao2015micro} for details. Since neither the working assumptions nor the inputs to the sample size formula specify the error distribution, we consider five distributions for the outcomes, including independent normal, correlated normal with different correlation structures,  independent $t$-distribution with three degrees of freedom (heavy tailed) and independent (centered) exponential distribution with rate parameter equal to 1 (skewed). The simulation results are provided in Table \ref{worktrue}.  In general, these results show that the power performance is still quite robust to different error distributions when the sample size equal to 10, as in the case of relatively large sample size demonstrated in \citet{liao2015micro}.  

\begin{sidewaystable}
	\begin{tabular}{ccccc||ccccccccc}
		\multicolumn{1}{p{0.6cm}}{\centering \ \\ \ \\$D$} &	  
		\multicolumn{1}{p{0.6cm}}{\centering \ \\ \ \\$K$} &	  
		\multicolumn{1}{p{0.6cm}}{\centering \ \\ \ \\$Z$} &	  
		\multicolumn{1}{p{0.6cm}}{\centering \ \\ \ \\ $\bar d$} &	    
		\multicolumn{1}{p{2cm}}{\centering \ \\Estimated\\ Power} &
		\multicolumn{1}{p{1.3cm}}{\centering \ \\ i.i.d.  \\ Normal} &
		\multicolumn{1}{p{1.3cm}}{\centering \ \\ i.i.d.  \\ t dist.} &
		\multicolumn{1}{p{1.3cm}}{\centering \ \\ i.i.d.  \\ Exp.dist.} &
		\multicolumn{1}{p{1.3cm}}{\centering \ \\ AR \\ (-0.8)} &
		\multicolumn{1}{p{1.3cm}}{\centering \ \\ AR \\ (-0.5)} &
		\multicolumn{1}{p{1.3cm}}{\centering \ \\ AR \\ (0.5)} &
		\multicolumn{1}{p{1.3cm}}{\centering \ \\ AR \\ (0.8)}	&		
		\multicolumn{1}{p{1.3cm}}{\centering \ \\ CSblock \\  (0.5)} &
		\multicolumn{1}{p{1.3cm}}{\centering \ \\ CSblock \\  (0.8)}  
		
		\\ \hline
		100 &   5 &    0 &    0.12 &  0.839 &    0.818 & 0.806 &   0.820 &   0.813 &   0.801 &  0.847 &  0.817 &     0.926 &     0.936 \\
		100 &   5 &    1 &    0.15 &  0.914 &    0.891 & 0.906 &   0.892 &   0.885 &   0.896 &  0.871 &  0.879 &     0.961 &     0.977 \\
		100 &   5 &    2 &    0.20 &  0.907 &    0.854 & 0.861 &   0.862 &   0.859 &   0.859 &  0.862 &  0.854 &     0.929 &     0.947 \\
		50 &  10 &    0 &    0.12 &  0.839 &    0.816 & 0.814 &   0.806 &   0.818 &   0.820 &  0.818 &  0.826 &     0.867 &     0.888 \\
		50 &  10 &    1 &    0.15 &  0.915 &    0.886 & 0.882 &   0.884 &   0.900 &   0.906 &  0.904 &  0.897 &     0.932 &     0.936 \\
		50 &  10 &    2 &    0.20 &  0.907 &    0.876 & 0.868 &   0.848 &   0.856 &   0.823 &  0.847 &  0.861 &     0.899 &     0.928 \\
		25 &  25 &    0 &    0.12 &  0.908 &    0.891 & 0.884 &   0.903 &   0.906 &   0.888 &  0.899 &  0.896 &     0.906 &     0.916 \\
		25 &  25 &    1 &    0.15 &  0.963 &    0.945 & 0.948 &   0.956 &   0.940 &   0.938 &  0.950 &  0.961 &     0.962 &     0.955 \\
		25 &  25 &    2 &    0.20 &  0.955 &    0.928 & 0.932 &   0.936 &   0.943 &   0.930 &  0.942 &  0.921 &     0.940 &     0.941 \\
		10 &  50 &    0 &    0.12 &  0.839 &    0.806 & 0.809 &   0.816 &   0.823 &   0.815 &  0.802 &  0.794 &     0.824 &     0.842 \\
		10 &  50 &    1 &    0.15 &  0.926 &    0.903 & 0.893 &   0.902 &   0.887 &   0.916 &  0.892 &  0.890 &     0.908 &     0.911 \\
		10 &  50 &    2 &    0.20 &  0.912 &    0.868 & 0.850 &   0.879 &   0.865 &   0.867 &  0.850 &  0.859 &     0.889 &     0.881 \\ \hline
	\end{tabular}
	
	\caption{Simulation results when working assumptions are true.  
		$D$ = Number of Days. $K$ = Number of decision times per day. $Z$ refers to the parameterization of the treatment effect in both the sample size model and simulation: 0 = Constant, 1 = Linear, and 2 = Quadratic.  $\bar d$ is the average standardized treatment effect. In all cases, the initial effect is 0, the treatment effects are identical within same day and the maximal effect is reached midway through the study. The underlying true effects in the generative model are the same as in the sample size model.  The expected availability is assumed to be constant throughout the study and equal to 0.7. 
		For the error distributions: the $t$ distribution is used with 3 degrees of freedom; the rate parameter in the exponential distribution is 1; AR($\rho$) and CSblock($\rho$) are the correlated normal distributions with correlation structure $\Sigma = (\Sigma_{ij})$ satisfying $\Sigma_{ij} = \rho^{\abs{i-j}}$, and $\Sigma_{ij} = \rho$ for $i \neq j$ in the same day, otherwise 0, respectively. Results are based on 1,000 replications.
		%
	}
	\label{worktrue}
\end{sidewaystable}

\begin{figure}
	\centering
	\begin{subfigure}[t] {.3\textwidth}
		\includegraphics[width = \textwidth]{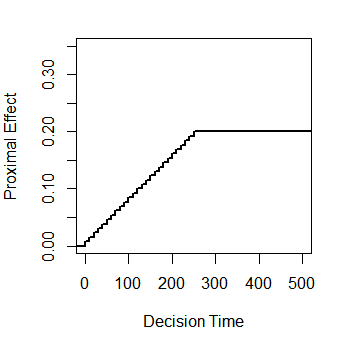}
		\caption{Trend 1: Maintained effect}
	\end{subfigure}
	\begin{subfigure}[t]{.3\textwidth}
		\includegraphics[width = \textwidth]{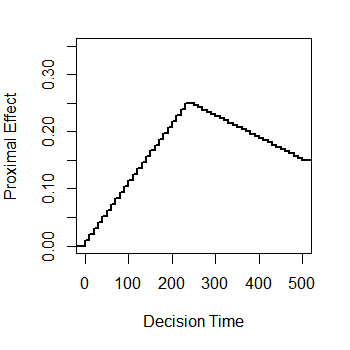}
		\caption{Trend 2: Slightly degraded effect}
	\end{subfigure}
	\begin{subfigure}[t]{.3\textwidth}
		\includegraphics[width = \textwidth]{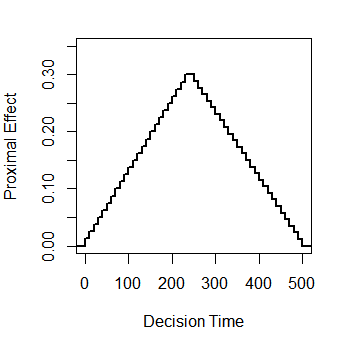}
		\caption{Trend 3: severely degraded effect}
	\end{subfigure}
	\caption{Proximal Treatment Effects $\{\beta(t)\}_{t=1}^T$: representing maintained (a), slightly degraded (b), or severely degraded (c) time-varying treatment effects. The horizontal axis is the decision time point. The vertical axis is the standardized treatment effect. }
	\label{ShapeOfBeta}
\end{figure}

Secondly, we consider the case in which the working assumptions proposed in \cite{liao2015micro} are not satisfied. In particular, we consider two cases. 
In the first case, the time-varying pattern of underlying true proximal effects is different from the user-provided pattern. For example, the user might input a linear pattern for the treatment effect, but the true effects are quadratic. Instead the vector of standardized effect, $d$ used in the sample size formula corresponds to the projection of $d(t)$, that is,
$d = (\sum_{t=1}^{T} E[I_t] Z_tZ_t^T)^{-1} \sum_{t=1}^{T}(E[I_t]Z_td(t))$; see more details in \citet{liao2015micro}. We consider three different patterns of treatment effect which cannot be represented as constant, linear and quadratic form, but can be sufficiently well approximated; see Figure \ref{ShapeOfBeta}. Results are provided in Table \ref{betat_wrong}. As opposed to the case when sample size is relatively large, the simulated power results when $N = 10$ are slightly smaller than the power estimated by MRT-SS Calculator, roughly below $0.05$. 

\begin{figure}
	\centering
	\begin{subfigure}[t] {.3\textwidth}
		\includegraphics[width = \textwidth]{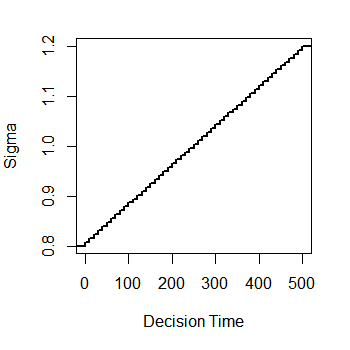}
		\caption{Trend 1: Linearly increasing}
		\label{fig:trendofsigma-lin-inc}
	\end{subfigure}
	\begin{subfigure}[t]{.3\textwidth}
		\includegraphics[width = \textwidth]{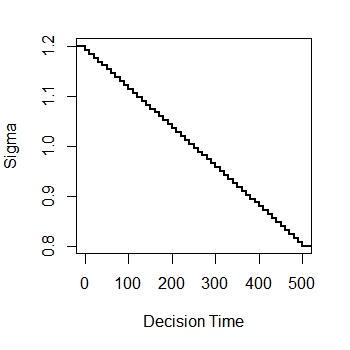}
		\caption{Trend 2: Linearly decreasing}
		\label{fig:trendofsigma-lin-dec}
	\end{subfigure}
	\begin{subfigure}[t]{.3\textwidth}
		\includegraphics[width = \textwidth]{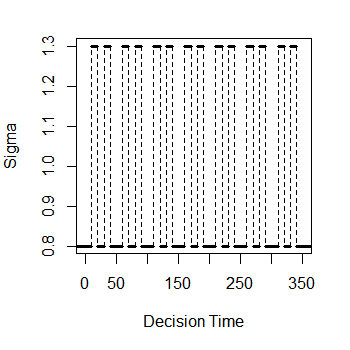}
		\caption{Trend 3: Jump discontinuous}
	\end{subfigure}
	\caption{Trends of $\bar \sigma_t$. For all trends, $\bar \sigma_t^2$ is scaled so that $(1/T)\sum_{t=1}^{T}\bar \sigma_t^2 = 1$. The horizontal axis is the decision time point. The vertical axis is the average conditional variance $\bar \sigma_t^2$.}
	\label{TrendofSigma}
\end{figure}

In the second case, we investigate the performance when the conditional variance of the outcome at decision time $t$, e.g. $\operatorname{Var}[Y_{t+1}|I_t = 1, A_t] = A_t\sigma_{1t}^2 + (1 - A_t)\sigma_{0t}^2$, is time-varying and depends on the treatment variable $A_t$. It was reported in \cite{liao2015micro} that when sample sizes are relatively large, power might decrease slightly depending on the choice of $\sigma_{0t}$ and $\sigma_{1t}$. Here, we investigate whether robustness is maintained in small sample sizes (e.g. $N = 10$). We consider three time-varying trends for the average conditional variance $\bar \sigma_{t}^2 = \rho \sigma_{1t}^2 + (1-\rho)\sigma_{0t}^2$, together with different ratio of $\sigma_{1t}$ and $\sigma_{0t}$; see Figure \ref{TrendofSigma}.	The simulation results are given in Table \ref{ratio_wrong}. It can be seen that the ratio factor $\sigma_{1t}/\sigma_{0t}$ has almost no impact on the simulated power. Large variation in $\bar \sigma_t$, e.g. trend 3 in Figure \ref{TrendofSigma}, reduces the power in all cases. This is also true when sample size is relatively large: the reduction in power is similar, on average 0.05. When treatment effects are constant, or quadratic with maximal effect midway through the study, either decreasing or increasing $\bar \sigma_t$ does not affect power substantially. When treatment effects are linear, an increasing trend (Figure~\ref{fig:trendofsigma-lin-inc}) lowers the power, while a decreasing trend (Figure~\ref{fig:trendofsigma-lin-dec}) improves the power.  

\begin{table}
	\centering
	\begin{tabular}{ccccccc}
		
		\multicolumn{1}{p{1cm}}{\centering  \ \\$D$} &	  
		\multicolumn{1}{p{1cm}}{\centering  \ \\$K$} &	  
		\multicolumn{1}{p{1cm}}{\centering  \ \\$Z$} &	  
		\multicolumn{1}{p{1cm}}{\centering  \ \\$\beta(t)$} &	 			  		
		\multicolumn{1}{p{1cm}}{\centering \ \\ $\bar d$} &	    
		\multicolumn{1}{p{2cm}}{\centering Estimated\\ Power} &
		\multicolumn{1}{p{2cm}}{\centering Simulated \\ Power} \\ \hline			  		
		
		100 &           5 &    2 &     (a) &       0.20 &  0.905 & 0.828 \\
		100 &           5 &    2 &     (b) &       0.20 &  0.901 & 0.833 \\
		100 &           5 &    2 &     (c) &       0.20 &  0.931 & 0.897 \\
		50 &          10 &    2 &     (a) &       0.20 &  0.906 & 0.867 \\
		50 &          10 &    2 &     (b) &       0.20 &  0.903 & 0.851 \\
		50 &          10 &    2 &     (c) &       0.20 &  0.933 & 0.898 \\
		25 &          25 &    2 &     (a) &       0.20 &  0.957 & 0.927 \\
		25 &          25 &    2 &     (b) &       0.20 &  0.960 & 0.942 \\
		25 &          25 &    2 &     (c) &       0.20 &  0.977 & 0.972 \\
		10 &          50 &    2 &     (a) &       0.20 &  0.920 & 0.875 \\
		10 &          50 &    2 &     (b) &       0.20 &  0.917 & 0.872 \\
		10 &          50 &    2 &     (c) &       0.20 &  0.952 & 0.933 \\ \midrule
		100 &           5 &    1 &     (a) &       0.15 &  0.871 & 0.822 \\
		100 &           5 &    1 &     (b) &       0.15 &  0.841 & 0.787 \\
		100 &           5 &    1 &     (c) &       0.15 &  0.820 & 0.758 \\
		50 &          10 &    1 &     (a) &       0.15 &  0.873 & 0.835 \\
		50 &          10 &    1 &     (b) &       0.15 &  0.842 & 0.808 \\
		50 &          10 &    1 &     (c) &       0.15 &  0.820 & 0.755 \\
		25 &          25 &    1 &     (a) &       0.15 &  0.923 & 0.896 \\
		25 &          25 &    1 &     (b) &       0.15 &  0.904 & 0.857 \\
		25 &          25 &    1 &     (c) &       0.15 &  0.896 & 0.868 \\
		10 &          50 &    1 &     (a) &       0.15 &  0.889 & 0.877 \\
		10 &          50 &    1 &     (b) &       0.15 &  0.853 & 0.814 \\
		10 &          50 &    1 &     (c) &       0.15 &  0.822 & 0.768 \\ \midrule
		100 &           5 &    0 &     (a) &       0.12 &  0.839 & 0.825 \\
		100 &           5 &    0 &     (b) &       0.12 &  0.839 & 0.818 \\
		100 &           5 &    0 &     (c) &       0.12 &  0.839 & 0.828 \\
		50 &          10 &    0 &     (a) &       0.12 &  0.839 & 0.792 \\
		50 &          10 &    0 &     (b) &       0.12 &  0.839 & 0.821 \\
		50 &          10 &    0 &     (c) &       0.12 &  0.839 & 0.774 \\
		25 &          25 &    0 &     (a) &       0.12 &  0.908 & 0.890 \\
		25 &          25 &    0 &     (b) &       0.12 &  0.908 & 0.886 \\
		25 &          25 &    0 &     (c) &       0.12 &  0.908 & 0.893 \\
		10 &          50 &    0 &     (a) &       0.12 &  0.839 & 0.819 \\
		10 &          50 &    0 &     (b) &       0.12 &  0.839 & 0.833 \\
		10 &          50 &    0 &     (c) &       0.12 &  0.839 & 0.826 \\ \hline
	\end{tabular}
	\caption{Simulation result when treatment effect is wrongly specified.  
		$D$ = Number of Days. $K$ = Number of decision times per day. $Z$ refers to the parameterization of the treatment effect in both sample size model and simulation; 0 = Constant, 1 = Linear, and 2 = Quadratic.  $\beta(t)$ is the underlying true treatment effect in the generative model; entries correspond to subfigures of Figure~\ref{ShapeOfBeta}.  $\bar d$ is the average of standardized treatment effect. The expected availability is assumed to be constant throughout the study and equal to 0.7. The error distribution in the generative model is i.i.d. normal. Results are based on 1,000 replications.}
	\label{betat_wrong}	
\end{table}

\begin{sidewaystable}
	\centering

	\begin{tabular}{ccccc||ccc|ccc|ccc}
		&     &     &          & Estimated & \multicolumn{3}{c|}{ratio = 0.8} & \multicolumn{3}{c|}{ratio = 1.0} & \multicolumn{3}{c}{ratio = 1.2} \\
		$D$ & $K$ & $Z$ & $\bar d$ &   power   & Trend 1 & Trend 2 &   Trend 3    & Trend 1 & Trend 2 & Trend 3  & Trend 1 & Trend 2 &  Trend 3  \\ \hline
		100 &           5 &    2 &       0.20 &     0.907 &             0.866 &             0.864 &             0.828 &           0.864 &           0.859 &           0.819 &             0.848 &             0.874 &             0.834 \\
		50 &          10 &    2 &       0.20 &     0.907 &             0.843 &             0.863 &             0.827 &           0.851 &           0.850 &           0.830 &             0.847 &             0.854 &             0.828 \\
		25 &          25 &    2 &       0.20 &     0.955 &             0.932 &             0.940 &             0.917 &           0.916 &           0.938 &           0.908 &             0.935 &             0.935 &             0.901 \\
		10 &          50 &    2 &       0.20 &     0.912 &             0.845 &             0.891 &             0.830 &           0.852 &           0.900 &           0.849 &             0.846 &             0.883 &             0.849 \\
		100 &           5 &    1 &       0.15 &     0.914 &             0.830 &             0.948 &             0.878 &           0.811 &           0.942 &           0.849 &             0.831 &             0.940 &             0.875 \\
		50 &          10 &    1 &       0.15 &     0.915 &             0.823 &             0.941 &             0.878 &           0.811 &           0.950 &           0.854 &             0.800 &             0.959 &             0.871 \\
		25 &          25 &    1 &       0.15 &     0.963 &             0.887 &             0.991 &             0.946 &           0.898 &           0.983 &           0.938 &             0.920 &             0.990 &             0.935 \\
		10 &          50 &    1 &       0.15 &     0.926 &             0.802 &             0.962 &             0.892 &           0.803 &           0.959 &           0.887 &             0.847 &             0.960 &             0.887 \\
		100 &           5 &    0 &       0.12 &     0.839 &             0.803 &             0.816 &             0.774 &           0.820 &           0.787 &           0.789 &             0.796 &             0.832 &             0.781 \\
		50 &          10 &    0 &       0.12 &     0.839 &             0.815 &             0.813 &             0.793 &           0.832 &           0.810 &           0.758 &             0.798 &             0.805 &             0.778 \\
		25 &          25 &    0 &       0.12 &     0.908 &             0.895 &             0.879 &             0.888 &           0.888 &           0.876 &           0.895 &             0.890 &             0.893 &             0.880 \\
		10 &          50 &    0 &       0.12 &     0.839 &             0.822 &             0.825 &             0.785 &           0.807 &           0.822 &           0.777 &             0.827 &             0.814 &             0.787 \\ \hline
	\end{tabular}
	\caption{Simulation result when the conditional variance of the outcome is time-varying and depends on the treatment variable.
		$D$ = Number of Days. $K$ = Number of decision times per day. $Z$ refers to the parameterization of the treatment effect in both sample size model and simulation; 0 = Constant, 1 = Linear, and 2 = Quadratic.  $\bar d$ is the average of standardized treatment effect. In all cases, the initial effect is 0, the treatment effects are identical within same day and attains the maximal effect midway through the study.  The underlying true effects in the generative model are same as in the sample size model.  The expected availability is assumed to be constant throughout the study and equal to 0.7. The ratio is defined as $\sigma_{1t}/\sigma_{0t}$ and is assumed constant. The trend refers to three time-varying patterns of the average conditional variance $\{\bar \sigma_t^2\}_{t=1}^T$; see Figure \ref{TrendofSigma}. Results are based on 1,000 replications.
		%
	}
	\label{ratio_wrong}
\end{sidewaystable}

\end{document}